\documentclass[aps,pra,floatfix,twocolumn,floatfix]{revtex4}

\begin{document}

\title{Calculation of the spectra of the superheavy element Z=112}

\author{T. H. Dinh, V. A. Dzuba, and V. V. Flambaum}

\affiliation{School of Physics, University of New South Wales,
        Sydney, 2052, Australia}
\date{\today}

\begin{abstract}

Accurate {\em ab initio} calculations of the energy levels of the superheavy
elements Z=112 are presented. Relativistic Hartree-Fock and configuration
interaction methods are combined with the many-body perturbation theory
to construct the many-electron wave function for valence electrons and
to include core-valence correlations. Two different approaches 
in which the element is treated as a system with two or twelve
external electrons above closed shells are used and compared.
Similar calculations for mercury are used to control the accuracy of the 
calculations. The results are compared with other calculations. 

\end{abstract}


\maketitle

\section{Introduction}

Study of the superheavy elements with nuclear charge $Z>100$ is an important 
area of research motivated by the search for the {\em island of stability}
(see, e.g.~\cite{rev1,rev2,rev3}).
Synthesis and investigation of superheavy elements are conducted at
leading nuclear-physics laboratories in Dubna, Berkeley, Darmstadt and
others. Elements with nuclear charge up to $Z$=118 have been 
synthesized~\cite{dubna}. 

The study of the element Uub ($Z$=112) is an important part of this research.
Since it was synthesized in Darmstadt in 1996~\cite{Hofmann} there were
numerous works discussing its production, nuclear and chemical
properties, etc. (see, e.g. Refs.~\cite{r112a,r112b,r112c,r112d,r112e,
Mosyagin,Nature112} and references therein). In contrast, only very few works
devoted to the study of electron structure and optical spectrum of the
element. Eliav {\em et al}~\cite{Kaldor} calculated ionization potential of
neutral Uub and few low energy levels of Uub$^+$ and Uub$^{2+}$. More detailed
study of neutral Uub were recently reported in
Refs.~\cite{Yu,JiGuang}. Quantum electrodynamic corrections (QED) for the Uub
element were studied in Ref.~\cite{QED112}. 

In present paper we try to address the shortage of data on the electron
structure and energy spectrum of the Uub element by calculating its energy
levels. Element 112 has electron structure similar to those of
mercury. Therefore, we use the calculations for mercury as a test of the
calculations and as a guide for their accuracy. Most of lower states of both
atoms can be considered as states with two valence electrons above closed
shells. We use the combined configuration interaction and many-body
perturbation theory method (CI+MBPT) \cite{CI+MBPT,DJ} to perform calculations 
for such states. This method has been successfully used for many different
atoms~\cite{VN,Ge,Ba1,Ba2,core} including the superheavy element with 
$Z$=120~\cite{E120}.

There are also states in mercury and element 112 with excitations from the
$5d$ or $6d$ subshell. They cannot be considered as two-electron states and 
in this case we use a version of the configuration
interaction (CI) technique with has been developed for atoms with open $d$ or
$f$ shells~\cite{FeI,DyI}. Some states are covered by both methods which is
another test of the accuracy of the calculations. We also compare our results
with the calculations of Li {\em et al} in Ref.~\cite{JiGuang}.

\section{Method of calculations and results for mercury}

Many states of mercury and element 112 (E112) can be considered as having 
two valence electrons above closed shells. The uppermost core subshell is 
the $5d^{10}$ subshell for mercury and the $6d^{10}$ subshell for E112. 
However, it is well known that mercury also has states of discrete spectrum 
which have one electron excited from the $5d^{10}$ subshell~\cite{Moore}. 
The lowest such state, the $5d^96s^26p \ ^3$P$^o_2$ state is obviously due to the 
$5d_{5/2} \rightarrow 6p_{1/2}$ excitation. Its energy is 68886.60 cm$^{-1}$
which is roughly double of the minimal excitation energy (see 
Table~\ref{Hgtab}). It is clear that the $6d_{5/2} \rightarrow 7p_{1/2}$ 
excitations in the E112 superheavy element must be even easier due to larger
fine structure. Indeed, with fine structure increasing the $6d_{5/2}$ and
$7p_{1/2}$ states move towards each other on the energy scale. The $6d_{5/2}$ 
state goes up while the $7p_{1/2}$ state goes down. This means that one should 
expect of having even more states with excitations for the $6d^{10}$ subshell 
in the discrete spectrum of E112 than those found in mercury. And these
states are expected to have lower energies.

The presence of the states with the $d-p$ excitations from the core is a 
serious complication for the calculations. The two-valence-electrons atoms
like Ba, Ra, E120~\cite{DJ,Ba1,Ba2,E120} can be treated very accurately by
means of the configuration interaction (CI) technique combined with the
many-body perturbation theory (MBPT) (the CI+MBPT method~\cite{CI+MBPT}).
In this method the CI technique is used to construct the two-electron
wave function and to include correlations between two valence electrons to
all orders via matrix diagonalization. The MBPT is used to include the
core-valence correlations. This method does include the core-valence
excitations but in an approximate way, using the lowest order perturbation 
theory. This might be not very accurate in the case when states with the
core-valence excitations are in the discrete spectrum, like in mercury and
E112.

The aim of present work is to predict the spectrum of the E112 superheavy
element. Since it has both types of states, with and without excitations 
from the $6d^{10}$ subshell, we use two different methods of calculations.
One is the CI+MBPT method for two valence 
electrons~\cite{CI+MBPT,DJ,Ba1,Ba2,E120} (method A) and another is the
CI method for twelve electrons~\cite{FeI,DyI} (method B). We demonstrate that
unless a two-electron state happens to be very close in energy to a state
with the excitation of the $d$-electron from the core the CI+MBPT method
gives remarkably accurate results. The twelve-valence-electrons method B
is used to find positions of the states with the excitations from the core.

\subsection{CI for two electrons: method A}

Here we use the CI+MBPT method developed in our earlier works~\cite{CI+MBPT,
DJ,Ba1,Ba2}.
The calculations are done in the $V^{N-2}$ approximation~\cite{VN} which means 
that initial Hartree-Fock procedure is done for a double ionized ion, with
two valence electrons removed. 

The effective CI Hamiltonian for a neutral two-electron atom is the sum
of two single-electron Hamiltonians plus an operator representing
interaction between valence electrons:
\begin{equation}
  \hat H^{\rm eff} = \hat h_1(r_1) + \hat h_1(r_2) + \hat h_2(r_1,r_2).
\label{heff}
\end{equation}
The single-electron Hamiltonian for a valence electron has the form
\begin{equation}
  \hat h_1 = h_0 + \hat \Sigma_1,
\label{h1}
\end{equation}
where $h_0$ is the relativistic Hartree-Fock Hamiltonian: 
\begin{equation}
  \hat h_0 = c \mathbf{\alpha p} + (\beta -1)mc^2 - \frac{Ze^2}{r} + V^{N-2},
\label{h0}
\end{equation}
and $\hat \Sigma_1$ is the correlation potential operator which represents 
correlation interaction of a valence electron with the core. 

Interaction between valence electrons is the sum of Coulomb interaction
and correlation correction operator $\hat \Sigma_2$:
\begin{equation}
  \hat h_2 = \frac{e^2}{|\mathbf{r_1 - r_2}|} + \hat \Sigma_2(r_1,r_2),
\label{h2}
\end{equation}
$\hat \Sigma_2$ represents screening of Coulomb interaction between 
valence electrons by core electrons. We use the second-order MBPT to calculate 
correlation operators $\hat \Sigma_1$ and $\hat \Sigma_2$.
The details can be found in our earlier works~\cite{CI+MBPT,DJ,Ba1,Ba2,E120}.

Two-electron wave function for the valence electrons $\Psi$ has a form of 
expansion over single-determinant wave functions
\begin{equation}
  \Psi = \sum_i c_i \Phi_i(r_1,r_2).
\label{psi}
\end{equation}
$\Phi_i$ are constructed from the single-electron valence basis 
states calculated in the $V^{N-2}$ potential
\begin{equation}
 \Phi_i(r_1,r_2) = \frac{1}{\sqrt{2}}(\psi_a(r_1)\psi_b(r_2)-\psi_b(r_1)\psi_a(r_2)).
\label{psiab}
\end{equation}
The coefficients $c_i$ as well as two-electron energies are found by
solving the matrix eigenvalue problem
\begin{equation}
  (H^{\rm eff} - E)X = 0,
\label{Schr}
\end{equation}
where $H^{\rm eff}_{ij} = \langle \Phi_i | \hat H^{\rm eff} | \Phi_j \rangle$ and
$X = \{c_1,c_2, \dots , c_n \}$.

The results of calculations for Hg with method A will be discussed in 
section~\ref{Hg-res}.


\subsection{CI for twelve electrons: method B}

The method used in this section has been developed in our earlier
works~\cite{FeI,DyI}.
As for the case of two valence electrons the method is based on the CI
technique. The main differences between method A and method B are in the 
choice of the basis and in the treatment of the core-valence correlations
(see below). In general, the method B is less accurate than method A.
However, its strong feature is the ability to deal with large number
of valence electrons.

The effective Hamiltonian for valence electrons has the form
\begin{equation}
  \hat H^{\rm eff} = \sum_{i=1}^{12}\hat h_{1i} + 
  \sum_{i < j}^{12} e^2/r_{ij},
\label{heffb}
\end{equation}
$\hat h_1(r_i)$ is the one-electron part of the Hamiltonian
\begin{equation}
  \hat h_1 = c \mathbf{\alpha \cdot p} + (\beta -1)mc^2 - \frac{Ze^2}{r} 
 + V_{core} + \delta V.
\label{h1b}
\end{equation}
Here $\mathbf{\alpha}$ and $\beta$ are Dirac matrixes, $V_{core}$ is the
Hartree-Fock potential due to core electrons and $\delta V$
is the term which simulates the effect of the correlations between core
and valence electrons. It is often called {\em polarization potential} and
has the form
\begin{equation}
  \delta V = - \frac{\alpha_p}{2(r^4+a^4)}.
\label{dV}
\end{equation}
Here $\alpha_p$ is polarization of the core and $a$ is a cut-off parameter
(we use $a = a_B$).

The differences between the Hamiltonian (\ref{heff}) in the previous section
and the Hamiltonian (\ref{heffb}) are: (a) the $5d$ electrons are treated as
core electrons in (\ref{heff}) and their contribution is included into the
potential $V^{N-2}$ while the $5d$ electrons are treated as valence electrons
in (\ref{heffb}) and their contribution is not included into the potential 
$V_{core}$; (b) the $\hat \Sigma_2$ operator is not included in (\ref{heffb});
the $\hat \Sigma_1$ operator in (\ref{heff}) is replaced by a less accurate
polarization potential $\delta V$ in (\ref{heffb}).

\begin{table}
\caption{Even and odd configurations of Hg and E112 and effective core
polarizability $\alpha_p$ (a.u.) used in the calculations.}
\label{sets}
\begin{ruledtabular}
  \begin{tabular}{l l l l l}
\multicolumn{1}{c}{Atom} &
\multicolumn{1}{c}{Set} &\multicolumn{1}{c}{Parity} & 
\multicolumn{1}{c}{Configuration} & 
\multicolumn{1}{c}{$\alpha_p$} \\
\hline
Hg   & 1 & Even &  $5d^{10}6s^2$ & 0.4 \\
     & 2 & Even &  $5d^{10}6p^2$ & 0.4 \\
     & 3 & Odd  &  $5d^{10}6s6p$ & 0.386 \\
     & 4 & Odd  &  $5d^96s^26p$  & 0.41 \\
E112 & 1 & Even &  $6d^{10}7s^2$ & 0.4 \\
     & 2 & Even &  $6d^{10}7p^2$ & 0.4 \\
     & 3 & Odd  &  $6d^{10}7s7p$ & 0.386 \\
     & 4 & Odd  &  $6d^97s^27p$  & 0.41 \\
\end{tabular}
\end{ruledtabular}
\end{table}

To construct the many-electron wave function for twelve valence electrons we
use the Hartree-Fock single-electron basis states which are found by the
self-consistent procedure performed independently for each configuration of
interest (see Refs.~\cite{FeI,DyI} for details). Table~\ref{sets} lists all
configurations of the valence electrons for mercury and E112 considered in
present work. The effective core polarizability parameter $\alpha_p$ is
treated as a fitting parameter. Its values for mercury are chosen to reproduce
the experimental data for energy levels of the corresponding
configurations. The same values are then used for the superheavy element E112.
The results for mercury will be discussed in next section.

\subsection{Results for mercury}

\label{Hg-res}

\begin{table*}
\caption{Energy levels of Hg (cm$^{-1}$)}
\begin{ruledtabular}
\begin{tabular}{llc l rrrr r}
\label{Hgtab}
        &      &     & & \multicolumn{4}{c}{Present work} & 
\multicolumn{1}{c}{Other\footnotemark[1]} \\
Config. & Term & $J$ & \multicolumn{1}{c}{$E_{\rm exp}$\footnotemark[2]} &
\multicolumn{1}{c}{$E_{A}$\footnotemark[3]} &
\multicolumn{1}{c}{$\Delta$\footnotemark[4]} &
\multicolumn{1}{c}{$E_{B}$\footnotemark[5]} &
\multicolumn{1}{c}{$\Delta$\footnotemark[4]} & \\
\hline
$5d^{10}6s^2$ & $^1$S     & 0 & \multicolumn{1}{r}{0.000} 
                                          &     0 &   0 &     0 &    0 &     0 \\
$5d^{10}6s6p$ & $^3$P$^o$ & 0 & 37645.080 & 37480 & 165 & 37763 & -699 & 38248 \\
              &           & 1 & 39412.300 & 39338 &  74 & 39442 &  -30 & 38441 \\
              &           & 2 & 44042.977 & 44287 &-244 & 42887 & 1156 & 49363 \\

$5d^{10}6s6p$ & $^1$P$^o$ & 1 & 54068.781 & 54263 &-194 & 54442 & -373 & 57402 \\
$5d^{10}6s7s$ & $^2$S     & 1 & 62350.456 & 62181 & 169 &       &      &       \\
$5d^{10}6s7s$ & $^1$S     & 0 & 63928.243 & 63681 & 247 &       &      &       \\
$5d^96s^26p$  & $^3$P$^o$ & 2 & 68886.60~ &       &     & 70287 &-1400 & 70139 \\
$5d^{10}6s7p$ & $^3$P$^o$ & 0 & 69516.66~ & 69223 & 294 &       &      &       \\
              &           & 1 & 69661.89~ & 69397 & 265 &       &      &       \\
              &           & 2 & 71207.51~ & 70106 &1102 &       &      &       \\
$5d^{10}6s7p$ & $^1$P$^o$ & 1 & 71295.15~ & 71213 &  82 &       &      &       \\

$5d^{10}6s6d$ & $^1$D     & 2 & 71333.182 & 71327 &   6 &       &      &       \\
$5d^{10}6s6d$ & $^3$D     & 1 & 71336.164 & 71345 &  -9 &       &      &       \\
              &           & 2 & 71396.220 & 71383 &  13 &       &      &       \\
              &           & 3 & 71431.311 & 71412 &  19 &       &      &       \\

$5d^96s^26p$  & $^3$D$^o$ & 3 & 73119.2~  &       &     & 71825 & 1294 & 71453 \\
$5d^96s^26p$  & $^1$P$^o$ & 1 & 78813~~~~ &       &     & 78174 &  639 & 79357 \\

\end{tabular}
\footnotetext[1]{J.G. Li {\em et al}, Ref.~\cite{JiGuang}}
\footnotetext[2]{Experiment, Ref.~\cite{Moore}}
\footnotetext[3]{Calculations with method A}
\footnotetext[4]{$\Delta = E_{\rm exp} - E_{A}$}
\footnotetext[5]{Calculations with method B}
\end{ruledtabular}
\end{table*}

Results for mercury are presented in Table~\ref{Hgtab}. Here experimental
energies are compared with the energies calculated within frameworks of
methods A and B which are described in previous section. Energy levels
of mercury were calculated by many authors before 
(see, e. g.~\cite{Yu,JiGuang,Glowacki}).
A review of these calculations goes beyond the scope of present work. 
In our case mercury serves only as a test of the calculations for the
superheavy elements Uub. Therefore we included in Table~\ref{Hgtab} 
the results of calculations of only one other group~\cite{JiGuang} who
also calculated the spectrum of Uub (see next section).

Method A gives very accurate results unless a state of interest happens
to be very close to another state with the same total momentum $J$ and
parity and which has a hole in the 5d shell. For example,
the largest deviation of the theory from experiment in method A is for
the $5d^{10}6s7p \ \ ^3$P$^o_2$ state which is close to the
$5d^96s^26p \ \ ^3$P$^o_2$ state. These states are strongly mixed,
however this mixing is included in a very approximate way in method A.
It treats an atom as a two-valence-electron system and excitations from 
the core are included only in the second-order of the MBPT in the $\hat \Sigma$
operator in the effective CI Hamiltonian. Note that this maximum deviation
(1102 cm$^{-1}$) is only 1.5\% of the energy.

Method B is less accurate, however it gives the positions of the energy
levels of the states with excitations from the 5d subshell which cannot
be obtained by method A.

The results of Ref.~\cite{JiGuang} are closer to our method B results.

\section{Results for Uub (Z=112)}

The results of calculations for the superheavy element Uub (Z=112) are
presented in Table~\ref{E112tab} together with the results of
Ref.~\cite{JiGuang}. We also present the $g$-factors Lande in the Table.
This includes the calculated $g$-factors as well as $g$-factors obtained from
analytical expressions in the $LS$ and $jj$ schemes. The $g$-factors are useful
for the identification of the states. The Uub is a superheavy element with
large relativistic effects. Therefore the $jj$ scheme works better for it than
the $LS$ one. However, the $LS$ scheme is also useful for the comparison with
mercury for which the $LS$ scheme is commonly used.

The $g$-factors in the $LS$ scheme are given by (non-relativistic notations)
\begin{eqnarray}
  && g_{NR}(J,L,S) = \label{gnr} \\
&&1 + \frac{J(J+1)-L(L+1)+S(S+1)}{2J(J+1)}, \nonumber
\end{eqnarray}
where $L$ is angular momentum of the atom, $S$ is its spin and $J$ is total
momentum ($\mathbf{J} = \mathbf{L} + \mathbf{S}$).
  
For the case of two electrons the $g$-factor in the $jj$ scheme is given by
\begin{eqnarray}
 && g_{jj}(J,j_1,j_2) = \label{gjj} \\
 && g_{NR}(j_1,l_1,1/2) \frac{J(J+1)-j_2(j_2+1)+j_1(j_1+1)}{2J(J+1)} \nonumber \\
 &&+g_{NR}(j_2,l_2,1/2) \frac{J(J+1)-j_1(j_1+1)+j_2(j_2+1)}{2J(J+1)}, \nonumber
\end{eqnarray}
where $j_1$ and $j_2$ are total momentum of each electron and $J$ is total
momentum of the atom ($\mathbf{J} = \mathbf{j_1} + \mathbf{j_2}$) and $g_{NR}$
is given by (\ref{gnr}). The formula (\ref{gjj}) also works for an electron
and a hole (e.g, the $(6d_{5/2}7p_{1/2})_2$ state).

The main difference in the spectra of mercury and Uub is due to larger fine
structure in the $6d$ subshell of Uub than in the $5d$ subshell of Hg. This
leads to easy excitation of the $6d_{5/2}$ electron and large number of the
states in the spectrum of Uub which correspond to the $6d^97s^27p$
configuration. According to calculations in Ref.~\cite{Kaldor} it also lead to
the change of the ground state configuration of Uub$^+$ as compared to the
Hg$^+$ ion. The ground state configuration of Uub$^+$ is shown to be the
$6d^97s^2$ configuration compared to the $5d^{10}6s$ ground state
configuration of Hg$^+$.
One should also note large negative relativistic correction for 7s energy
in Uub which makes this state to be more tightly bound than the 6s state of Hg.

The states of the $6d^97s^27p$ configuration are calculated with method B,
states of the $6d^{10}7s8s$ and $6d^{10}7s8p$ configurations are calculated
with method A and the states of the $6d^{10}7s7p$ configuration are calculated
with both methods. The results of both methods are in good agreement with each
other and in reasonable agreement with Ref.~\cite{JiGuang}. States of the
$6d^97s^27p$ configuration are well separated in energy from the states of the
same total momentum of the $6d^{10}7s7p$ and $6d^{10}7s8p$
configurations. This means that the mixing between these states is small and
should not affect the accuracy of the results. Judging by comparison with
mercury we expect the results for Uub to be accurate within few per cents.

Accurate calculations for superheavy elements should include Breit
interaction, quantum electrodynamic (QED) corrections and volume isotope
shift. However, as it has been demonstrated in our previous
works~\cite{E119,E120} even for atoms with Z=120 Breit and QED corrections are
relatively small and extrapolation of the error from lighter analogs of the
superheavy atoms is likely to produce more accurate results than the inclusion
of these small corrections. The accuracy of present calculations is lower than
that for Z=120 in Ref.~\cite{E119,E120} due to complex electron structure of the
Uub element. Therefore, these small corrections can be safely neglected on
the present level of accuracy. This is in agreement with the results or 
Ref.~\cite{QED112} in which QED corrections have been considered for E112
and found to contribute about 0.5\% to the ionization potential.

\begin{table*}
\caption{Calculated energies ($E$, cm$^{-1}$) and $g$-factors of ekamercury ($Z=112$).}
\begin{ruledtabular}
\begin{tabular}{lll cc c rcrcr}
\label{E112tab}
Config. & \multicolumn{2}{c}{Term} & \multicolumn{2}{c}{$g$-factors} & 
 \multicolumn{2}{c}{Method A} & 
\multicolumn{2}{c}{Method B}  & Other\footnotemark[1] \\
        &\multicolumn{1}{c}{$LS$} & \multicolumn{1}{c}{$j-j$} & $g_{\rm NR}$ & $g_{jj}$&  
\multicolumn{1}{c}{$E$} & \multicolumn{1}{c}{$g$}&
\multicolumn{1}{c}{$E$} & \multicolumn{1}{c}{$g$}&\multicolumn{1}{c}{$E$} \\
\hline

$6d^{10}7s^2$ & $^1$S$_0$  & $(7s_{1/2},7s_{1/2})_0$ & 0.00 & 0.00 &     0 & 0.00 &       0 & 0.00 & 0 \\  
                                                                                          
$6d^97s^27p$  & $^3$P$^o_2$ & $(6d_{5/2},7p_{1/2})^o_2$ & 1.50 & 1.29 &       &      & 35785 & 1.35 & 34150 \\
$6d^97s^27p$  & $^3$F$^o_3$ & $(6d_{5/2},7p_{1/2})^o_3$ & 1.08 & 1.11 &       &      & 38652 & 1.10 & 37642 \\
$6d^97s^27p$  & $^3$P$^o_4$ & $(6d_{5/2},7p_{3/2})^o_4$ & 1.25 & 1.25 &       &      & 56131 & 1.25 & 60366 \\

$6d^{10}7s7p$ & $^3$P$^o_0$ & $(7s_{1/2},7p_{1/2})^o_0$ & 0.00 & 0.00 & 51153 & 0.00 & 51212 & 0.00 & 48471 \\  
              & $^3$P$^o_1$ & $(7s_{1/2},7p_{1/2})^o_1$ & 1.50 & 1.33 & 55057 & 1.41 & 53144 & 1.33 & 52024 \\  
              & $^3$P$^o_2$ & $(7s_{1/2},7p_{3/2})^o_2$ & 1.50 & 1.50 & 73736 & 1.50 & 70416 & 1.49 & 76641 \\

$6d^97s^27p$  & $^3$D$^o_2$  & $(6d_{3/2},7p_{3/2})^o_2$ & 1.17 & 1.07 &       &      & 56960 & 1.12 & 60809 \\
$6d^97s^27p$  & $^3$P$^o_1$  & $(6d_{5/2},7p_{3/2})^o_1$ & 1.00 & 1.10 &       &      & 58260 & 1.15 & 64470 \\
$6d^97s^27p$  & $^3$P$^o_1$  & $(6d_{3/2},7p_{3/2})^o_1$ & 1.00 & 1.07 &       &      & 68673 & 1.00 & 73686 \\
                                                                                          
$6d^{10}7s7p$ & $^1$P$^o_1$  & $(7s_{1/2},7p_{3/2})^o_1$  & 1.00 & 1.17 & 79637 & 1.10 & 78697 & 1.04 & 85533 \\
                                                                                          
$6d^97s^27p$ &              & $(6d_{3/2},7p_{3/2})^o_0$  & 0.00 & 0.00 &       &      & 80442 & 0.00 & 82895 \\

$6d^{10}7s8s$ & $^2$S$_1$  & $(7s_{1/2},8s_{1/2})_1$  & 2.00 & 2.00 & 87785 & 2.00 &     &      & \\
$6d^{10}7s8s$ & $^1$S$_0$  & $(7s_{1/2},8s_{1/2})_0$  & 0.00 & 0.00 & 88861 & 0.00 &       &      & \\ 
                                                                                          
$6d^{10}7s8p$ & $^3$P$^o_0$ & $(7s_{1/2},8p_{1/2})^o_0$  & 0.00 & 0.00 & 95903 & 0.00 &       &      & \\
              & $^3$P$^o_1$ & $(7s_{1/2},8p_{1/2})^o_1$  & 1.50 & 1.33 & 95084 & 1.39 &       &      & \\
              & $^3$P$^o_2$ & $(7s_{1/2},8p_{3/2})^o_2$  & 1.50 & 1.50 & 97342 & 1.50 &       &      & \\ 
\end{tabular}
\footnotetext[1]{Reference.~\cite{JiGuang}}
\end{ruledtabular}
\end{table*}

\section{Conclusion}

We have calculated 17 lowest energy levels of the superheavy element Uub
(Z=112). Comparison with similar calculations for mercury indicate that the
accuracy of the calculations is within few per cents. The results can be used
in the study of the chemical and spectroscopic properties of the superheavy
element.

\section*{Acknowledgments}

The work was funded in part by the Australian Research Council.

\end{document}